\documentclass[final,5p,times,twocolumn]{elsarticle}
\usepackage[figuresright]{rotating}
\usepackage{bm}
\usepackage{dsfont}
\usepackage{amsmath}
\usepackage{amssymb}
\usepackage{color}
\usepackage{url}
\usepackage{hyperref}
\usepackage{algorithm,algorithmicx,algpseudocode}
\usepackage{bbm}

\def\dblone{\hbox{$1\hskip -1.2pt\vrule depth 0pt height 1.6ex width 0.7pt
                  \vrule depth 0pt height 0.3pt width 0.12em$}}

\begin{document}
\begin{frontmatter}
\title{Calculating Floquet states of large quantum systems:
\\ A parallelization strategy and its cluster implementation}

\author[tudm]{T.V.~Laptyeva}
\author[inf]{E.A.~Kozinov}
\author[inf]{I.B.~Meyerov}
\author[app]{M.V.~Ivanchenko}
\author[augs,app]{S.V.~Denisov\corref{cor1}}
\ead{sergey.denisov@physik.uni-augsburg.de}
\author[augs,app]{P.~H\"{a}nggi}
\cortext[cor1]{Corresponding author at: Institut f\"ur Physik, Universit\"at Augsburg,
Universit\"atsstra{\ss}e 1, D-86135 Augsburg, Germany}

\address[tudm]{Theory of Control and Dynamical Systems Department, Lobachevsky State University of Nizhny Novgorod, Russia}
\address[inf]{Mathematical Software and Supercomputing Technologies Department, Lobachevsky State University of Nizhny Novgorod, Russia}
\address[app]{Department of Applied Mathematics, Lobachevsky State University of Nizhny Novgorod, Russia}
\address[augs]{Institut f\"ur Physik, Universit\"at Augsburg, Universit\"atsstra{\ss}e 1, D-86135 Augsburg, Germany}

\begin{abstract}
We present a numerical approach to calculate non-equilibrium eigenstates
of  a periodically time-modulated quantum system.
The approach is based on the use of a chain of single-step
time-independent propagating operators. Each operator is time-specific and constructed by combining
the Magnus expansion of the time-dependent system Hamiltonian
with the Chebyshev expansion of an operator exponent. A construction of a unitary
matrix of the Floquet operator, which evolves a system state over the full modulation period,
is performed by propagating the identity matrix over the period.
The independence of
the evolutions of basis vectors makes the propagation stage suitable for implementation
on a parallel cluster. Once the propagation stage is completed, a routine diagonalization
of the  Floquet  matrix is performed. Finally, an additional propagation round, now with the
eigenvectors as the initial states, allows to resolve the time-dependence of the Floquet states
and calculate their characteristics. We demonstrate the accuracy and scalability of the algorithm by applying
it to calculate the Floquet states of two quantum models, namely (i) a synthesized random-matrix Hamiltonian
and (ii) a many-body Bose-Hubbard dimer, both of the size up to $10^4$ states.

\end{abstract}

\end{frontmatter}

\section{Introduction}\label{introduction}
Fast progress in manipulations with cold and ultra-cold atoms, quantum optics and
nanoscale fabrication techniques has brought quantum physics in touch with technology \cite{bloch,poot,asp}.
It is natural then
that computational quantum physics plays  an ever increasing role in  explaining and guiding
current experiments and suggesting new  \cite{device}.
From the computational point of view, the complete resolution of a coherent, i.e., an isolated from the environment,
quantum system means the solution of the eigenvalue problem for the system Hamiltonian $H$.
When the Hamiltonian is time-independent, this task can be executed by performing
full diagonalization of the Hamiltonian matrix
When the system becomes to large then
the size of the matrix  does not  allow
for the full diagonalization.  The task, however, could be
restricted to finding lowest energy eigenstate(s) which  can be accomplished
by using the Lanczos algorithm \cite{lan} or
more sophisticated tools,  such as the Density-Matrix Renormalization Group (DMRG) methods \cite{dmrg}.
In cases that the system is periodically modulated in time,
its Hamiltonian becomes a time-periodic matrix $H(t+T) = H(t+2\pi/\omega) = H(t)$. Then, the dynamics of the system is
governed by the set of  so termed \textit{Floquet eigenstates} \cite{shirley,sambe}. These states are not
eigenvectors of the Hamiltonian $H(t)$
but instead of the unitary Floquet operator
\begin{eqnarray}
U_T = \mathcal{T} \exp[-\frac{i}{\hbar} \int_0^T H(\tau) d\tau],
\label{prop}
\end{eqnarray}
where $\mathcal{T}$ is Dyson's time-ordering operator.
This   operator propagates the system over the one full period $\mathcal{T}$ of modulation,
while the corresponding time-periodic Floquet states (see below) at equal times $t=t'$ \cite{grifoni,kohler}
form a time-periodic  orthogonal basis spanning the system Hilbert space
and evolving under the action of the time-dependent Hamiltonian.
The structure of the unitary Floquet matrix, and thus properties
of the Floquet states, depend on the modulation
protocols and parameters. This is a key feature
of periodically driven quantum systems which makes them so attractive to the theoreticians and experimentalists
working in the filed of quantum optics, optomechanics and solid state physics \cite{grifoni,kohler,arimondo,bukov,eisert}.
Strong modulations can sculpt  a  set of non-equilibrium  eigenstates which may drastically differ
 from the states exhibited by the system in the unmodulated,  time-independent limit.
Modulations allow to grasp novel phenomena and effects
which are out of reach within  time-idependent  Hamiltonians; they can be used
to create topological insulators in semiconductor wells \cite{top}, synthesize
Majorana fermions in quantum wires \cite{majorana}, and engineer
gauge fields for spinless neutral atoms \cite{gauge}.

The calculation of Floquet states of a large quantum system constitutes a challenge.
The key step is a construction of the corresponding  unitary Floquet matrix, Eq.~~(\ref{prop}) (its final diagonalization
is such a routine as, for example, the diagonalization of stationary Hamiltonian matrices).
The most straightforward way to obtain $U_T$ is to numerically propagate
the identity matrix over the time period $T$. However, the propagation  with a time-dependent Hamiltonian operator
is an issue itself, to be addressed in the next section. There are two ways to do so.

The first option  is to use piecewise-constant
modulation functions. This allows  to reduce the computational task to the diagonalization of
time-independent  Hamiltonians, one for every time interval, and
the  expansion of eigenvectors of a preceding Hamiltonian in the basis of the consecutive one.
Such modulations were  used to investigate connections between integrability and
thermalization \cite{lazarides,lazarides2,rigol}, and to explore disorder-induced
localization \cite{abanin} in periodically driven many-body systems.
With respect to  the thermalization it was found that the modulations heat the system to the
infinite temperature so  that the system Floquet states are near uniformly smeared over the eigenbasis
of the  system in the absence of driving \cite{lazarides,lazarides2,rigol}.  An important
question that immediately arises is whether this is a universal
phenomenon or it is related to the  non-differentiability
of the modulation function (which property induces the presence of all  multiple frequencies $k\omega$,
$k = 1,2,...$, in the spectrum of
the modulations function).  Evidently, this question cannot be answered without going beyond the piecewise setup.
In addition, in the view of possible experimental realizations,  smooth continuous modulations are also
more preferable.

An alternative option is to expand the time-dependent Hamiltonian
into a Fourier series and, and then truncating it,
by keeping $2F+1$ harmonics $k\omega$,  $k= -F,...,0,...,F$ only, to
reduce the problem to the diagonalization  of a time-\textit{independent} super-Hamiltonian \cite{sambe,andre}.
This is a reliable method to obtain Floquet spectrum of a system of a
size up to a hundred of states.
For larger systems, this strategy leads to a computational problem: The size
of the super-Hamiltonian scales as $N \times (2F+1)$, where $N$ is the dimension of the system's Hilbert space.
Computational diagonalization efforts increase as $[N \times (2F+1)]^3$,
while the known diagonalization algorithms are poorly scalable.
For a system of the size $N=10^4$, already $F=50$ harmonics is too much;  a full
diagonalization of a  $10^6 \times 10^6$ matrix is unfeasible.
At the same time, this large number of harmonics is seemingly not enough to resolve faithfully
the Floquet spectrum of the system \footnote{The eigenvalue spectrum of the super-Hamiltonian can be resolved with
the accuracy $2\pi/M$ at best. This is not enough taking into account
that the actual mean spacing between the eigenvalues is $\pi/N$.}.


Therefore, in order to calculate the Floquet state of a system with $N \geqslant 10^3$ states,
the propagation stage has to be included into an algorithm. A propagation method should guarantee a high
accuracy with respect not only to the unitarity of the time evolution  but also with respect to the phases
of complex vectors. That is because  Floquet states appear as  superpositions of
basis vectors used to write system's Hamiltonian. Accumulated phase errors will destroy
the interference and lead to an incorrect set of Floquet states. As we show in Section \ref{results},
quantum interference effects, together with some facts from the quantum chaos theory, can be used to benchmark
the accuracy of an algorithm.

Because of the trade-off between the accuracy and system size, the time of sequential vector propagation
grows super-linearly with $N$. Faithful calculations of Floquet spectra of non-integrable  systems
(that are systems whose Hilbert space cannot be decomposed into several
non-interacting low-dimensional manifolds  \cite{nonint}),
with tens of thousands of states, can only be performed  with  scalable algorithms.

This paper presents an algorithm to calculate the Floquet spectra of strongly-modulated
quantum systems with $N \geqslant 10^4$ states
and its implementation  on a parallel supercomputer. The propagation part of the algorithm
is based on the combination of the Magnus expansion of time-dependent linear operators \cite{magnus}
and the Chebyshev expansion of operator exponents \cite{talezer}. This combination  has been proposed
in \cite{alverman}, where its particular numerical realization, implementing
a commutator-free Magnus scheme, was tested.
We illustrate the accuracy and scalability
of the algorithm by using two quantum models, with a synthesized random-matrix Hamiltonian and a many-body
non-integrable bosonic dimer. The size of model system is limited
by the diagonalization routine only, so the algorithm can be used to calculate Floquet states
of systems of the size up to $N \sim 50~000$ states.

The rest of the paper is organized as follows: Section~\ref{theory} outlines the
theoretical background and introduces
the Magnus and Chevyshev expansions;
Section~\ref{algorithm} describes the algorithm;
in Section~\ref{models} we introduce model systems,
apply the cluster implementation to calculate their
Floquet states in Section~\ref{realization},
and analyze the results in Section~\ref{results}.
Finally we summarized our findings and outline further perspectives in Section~\ref{summary}.

\section{Theoretical background}\label{theory}

\textit{Floquet states.} We consider  quantum systems whose dynamics is determined by the time-dependent Schr\"{o}dinger equation
\begin{eqnarray}
i\hbar \partial_t |\psi(t)\rangle = H(t)  |\psi(t)\rangle,
\label{SchrE}
\end{eqnarray}
where the Hamiltonian $H(t)$ denotes a time-periodic Hermitian operator, $H(t+T) = H(t)$.
We assume that the system evolves in
a finite-dimensional Hilbert space spanned by $N$ basis vectors.
The time evolution of the system is fully determined
by a unitary operator $U(t_0,t)$, being the solution
of the equation
\begin{eqnarray}
i\hbar \partial_t U(t_0,t) = H(t)  U(t_0,t),
\label{SchrU}
\end{eqnarray}
for the initial condition in the form of the identity matrix, $U(t_0,t_0) = \dblone$.  This provides the \textit{propagator}
of the system, i.e., a unitary operator which evolves any system state from a time $t_0$ to time $t_0+t$,
$U(t_0,t)|\psi(t_0)\rangle = |\psi(t_0+t)\rangle$. A time $t_0 \in [0,T]$  specifies the state of the
Hamiltonian operator at the initial time, when, for example, the driving was switched on. This starting time can be
absorbed into the Hamiltonian as a parameter, $H(t, t_0) = H(t+t_0)$ (the propagator 
$U(t_0,t)$ can be onbtained from $U(0,t)$
as $U(t_0,t) = U^{\dagger}(0,t_0)U(0,t+t_0)$), so for
later convenience, we set $t_0 = 0$ in Eq.~(\ref{SchrU}) and denote $U(0,t)$ by $U_t$.
Eigenvectors $\{| \varphi_{\mu} \rangle \}$ of the normal matrix $U_T$,
\begin{eqnarray}
U_T |\varphi_{\mu} \rangle = e^{-i\theta_{\mu}} |\varphi_{\mu} \rangle, ~~~ \mu = 1,\ldots,N,
\label{Floquet1}
\end{eqnarray}
form an orthonormal basis in the system Hilbert space. These vectors
could also be taken as  snapshots of time-dependent vectors $|\varphi_{\mu} (t) \rangle$
at the time instant $t=T$,
$U_t |\varphi_{\mu} (0) \rangle = e^{-i\epsilon_{\mu} t/\hbar}
|\varphi_{\mu} (t) \rangle$, with $\epsilon_{\mu} = \hbar \theta_{\mu}/T$.
The exponents $\epsilon_{\mu}$ have the dimension of energy and are termed \textit{quasienergies}.
Quasienergies can be determined up to multiples of $\hbar \omega$ so they are
conventionally restricted  to the interval $[-\hbar \omega/2, \hbar \omega/2]$.

By denoting  $ |\phi_{\mu} (t) \rangle = e^{-i\epsilon_{\mu} t/\hbar} |\varphi_{\mu} (t) \rangle$,
we end up with the set of time-periodic \textit{Floquet states} \cite{shirley,sambe,grifoni,kohler}
\begin{eqnarray}
|\phi_{\mu} (t+T) \rangle = |\phi_{\mu} (t) \rangle.
\label{Floquet2}
\end{eqnarray}

By knowing the Floquet spectrum of a system, $\{\epsilon_{\mu}, |\phi_{\mu} (t) \rangle\}$,  and the system initial
state $|\psi(0)\rangle$, one can calculate the state of the system at any instant of time $t > 0$,
\begin{eqnarray}
|\psi(t)\rangle  = \sum_n c_n e^{-i\epsilon_{\mu} t/\hbar} |\phi_{\mu} (t) \rangle, ~~~
c_n = \langle \psi(0)|\phi_{\mu} (0) \rangle.
\label{Floquet3}
\end{eqnarray}

\textit{Magnus expansion.}  The idea of the Magnus expansion \cite{magnus0} is to construct a time independent
Hamiltonian operator $\Omega(t_1,t_2)$, parameterized by the two times, $t_1$ and $t_2$, such that
\begin{eqnarray}
U(t_1,t_2) = \exp\left[-\frac{i}{\hbar}\Omega(t_1,t_2)\right].
\label{magnus1}
\end{eqnarray}

The operator is given by an infinite series involving nested commutators  \cite{magnus}
\begin{equation}
\begin{split}
\Omega(t_1,t_2) & = \int_{t_1}^{t_2} H(\tau_1) ~d\tau_1 + \\
& + \frac{1}{2} \int_{t_1}^{t_2} d\tau_1 \int_{t_1}^{\tau_1} [H(\tau_1), H(\tau_2)] ~d\tau_2 + \\
& \frac{1}{6} \int_{t_1}^{t_2} d\tau_1 \int_{t_1}^{\tau_1} d\tau_2
\int_{t_1}^{\tau_2} \big([H(\tau_1)[H(\tau_2), H(\tau_3)] + \\
& [H(\tau_3)[H(\tau_2), H(\tau_1)]\big) ~d\tau_3 + \ldots .
\label{magnus2}
\end{split}
\end{equation}

An implementation of the expansion (\ref{magnus2}) assumes a
truncation of the infinite series, summation of the finite series into an
operator $\Omega(t_1,t_2)$,  and use of the latter as the
propagator $U(t_1,t_2)$ \cite{magnus0}.  The Floquet operator $U_T$
can be approximated as a chain
\begin{align}
\begin{aligned}
U_T = U(0, t_1) & U(t_1, t_2) \ldots U(t_{M-1}, t_M) \approx \\
& \approx \ e^{-i \Omega(0,\: t_1)/\hbar}e^{-i \Omega(t_1,\: t_2)/\hbar} \ldots e^{-i \Omega(t_{M-1},\: t_M)/\hbar},
\label{magnus3}
\end{aligned}
\end{align}
where $t_k = k h = k T/M$, $k=0,...,M$. Since all terms on the rhs of Eq.~(\ref{magnus2}) are Hermitian,
the truncated operator $\Omega(t_1,t_2)$ is Hermitian, and
an approximation of any order preserves the unitary time evolution.
The truncated operator in the  form (\ref{magnus2}) is not very
suitable for computations. It is more convenient to approximate $\Omega(t_1,t_2)$
with lower-order commutator series, calculated by using values of $H(t_{1/2})$
at the midpoints $t_{1/2} = (t_{1} + t_2)/2$ (this is our choice,
see Section~\ref{algorithm} for more  details), or with
a commutator-free linear combination of $H(t_j)$, calculated at
different times $t_j \in [t_1,t_2]$ \cite{magnus,alverman}.

\textit{Chebyshev expansion.}  The exponentiation of an operator is a computationally expensive operation \cite{exp}.
In order to propagate vector $|\psi(t_1)\rangle$ to time $t_2$, the knowledge of the unitary operator
$\exp(-i \Omega(t_1,t_2)/\hbar)$ is redundant: we need only the result of its action on the vector,
$|\psi(t_2)\rangle = \exp(-i \Omega(t_1,t_2)/\hbar) \: |\psi(t_1)\rangle$.  This can be calculated
by implementing the Chebyshev polynomial expansion of the operator exponent, which is based on
a recursive iteration scheme \cite{talezer},
\begin{eqnarray}
|\psi_{l+1}(t_2)\rangle = -2 i \tilde{\Omega}(t_1,t_2) \: |\psi_{l}(t_2)\rangle + |\psi_{l-1}(t_2)\rangle
\label{chebyshev1}
\end{eqnarray}
with the initial conditions $|\psi_0(t_2)\rangle = |\psi(t_1)\rangle$
and $|\psi_1(t_2)\rangle = -i\tilde{\Omega}(t_1,t_2)\:|\psi_0(t_2)\rangle$.
Here $\Omega(t_1,t_2)$ is a shifted and rescaled operator,
\begin{eqnarray}
\tilde{\Omega}(t_1,t_2) = \frac{\Omega(t_1,t_2) - \dblone (\Delta E + E_{\rm min})}{\Delta E},
\label{newOmega}
\end{eqnarray}
which has all its eigenvalues restricted to the interval  $[-1,1]$ \cite{talezer}.
The spectral half-span $\Delta E = (E_{\rm max} - E_{\rm min})/2$ should  be
estimated from the extreme eigenvalues $E_{\rm min}$ and $E_{\rm max}$
of $\Omega(t_1,t_2)$ operator beforehand.

Finally, the new vector can be obtained as
\begin{eqnarray}
|\psi(t_2)\rangle = e^{-i \beta h / \hbar } \sum \limits_{l=0}^{L} a_l \: |\psi_l(t_2)\rangle,
\label{chebyshev2}
\end{eqnarray}
where $\beta = \Delta E + E_{\rm min}$ and $h = t_2 - t_1$. The expansion coefficients
$a_l = 2 J_l(R)$ and $a_0 = J_0(R)$, where $J_l(R)$ are the
Bessel functions of the first kind and $R = h \Delta E/ \hbar$.
The parameter $L$ sets the order of the Chebyshev approximation by truncating the series (\ref{chebyshev2}).
Strictly speaking, this scheme does not preserve the unitary time evolution. However, its
convergence with the increase of $L$
is  fast so that $L$ can be chosen such that the deviation from  unitarity is dominated
by the round-off error \cite{talezer}. We have found that it is enough to take $L < 100$
for $N \lesssim 10^4$ and the further increase of $L$ does not improve the accuracy of calculations.

\section{The algorithm} \label{algorithm}
We restrict the consideration to Hamiltonians of the form
\begin{eqnarray}
H(t) = H_0 + f(t)\cdot H_{\mathrm{mod}},~~~ f(t+T) = f(t),
\label{Ham1}
\end{eqnarray}
where $f(t)$ is a scalar function and $H_0$, $H_{\mathrm{mod}}$ are time-independent Hermitian operators.
Most of the currently used models, including the ones discussed in Section~\ref{models},
belongs to this class. Equation~(\ref{Ham1}) is the simplest nontrivial case of a general situation,
$H(t) = H_0 + \sum_s f_s(t)\cdot H^{(s)}_{\mathrm{mod}}$, with $s \leqslant N^2$. Our results
can be generalized to the case $s > 1$ in a straightforward manner.

Next we specify the method to approximate $\Omega(t_1,t_2)$. As we discussed  in the previous section,
there exist a variety of schemes \cite{magnus}. Our choice is conditioned by
the form of the Hamiltonian, Eq.~(\ref{Ham1}), and the intention to realize the algorithm
on a parallel cluster. More specific on the last point, we are not concerned about
the number of commutators needed to be calculated (and then stored)
as long as they are all time-\textit{in}depended and do not have to be recalculated in course of the propagation.
Here we use the midpoint approximation
of the Magnus expansion with three commutators \cite{magnus, blanes},
\begin{eqnarray}
\Omega = \alpha_1 + \frac{1}{12}\alpha_3 + \frac{1}{240}[-20\alpha_1 - \alpha_3 + C_1, \alpha_2 +C_2],
\label{magnusApp}
\end{eqnarray}
so that
$\Omega = \Omega + \mathcal{O}(h^7)$. Specifically,
\begin{eqnarray}
\alpha_j&=&\frac{h^j}{(j-1)!} \frac{d^{j-1}H(t_{1/2})}{dt^{j-1}},~~~~ C_1 = [\alpha_1, \alpha_2], \\
C_2&=&-\frac{1}{60}[\alpha_1, 2\alpha_3 + C_1]. \nonumber
\label{alphas}
\end{eqnarray}

The original formulation demands the calculation of $\alpha_j$ on every time step.
This task, for the specific choice given by Eq.~(\ref{Ham1}),
reduces to calculations of midpoint values of the scalar functions $f(t)$, $f'(t)$, and $f''(t)$. These values
have to be weighted with time-independent commutators of the forms $[H_0,H_{\mathrm{mod}}]$,
$[H_0,[H_0,H_{\mathrm{mod}}]]$,  etc. There are nine commutators for the chosen scheme, Eq.~(\ref{magnusApp}),
but they have to be calculated only once, when initiating the algorithm.

The choice of the  operational basis
to write operators $H_0$ and $H_{\mathrm{mod}}$ constitutes an important point.
We use of the eigenbasis of the operator $H_0$ \textit{without} going into the interaction picture
(see a relevant discussion in Ref.~\cite{alverman}). In this basis equation ~(\ref{Ham1}) assumes the form
\begin{eqnarray}
i\hbar \partial_t \: |\psi(t)\rangle = [{\rm diag} E_i + f(t) \cdot \tilde{H}_{\rm mod}] \: |\psi(t)\rangle,
\label{Ham2}
\end{eqnarray}
where  ${\rm diag} E_i$  is a diagonal matrix consisting of the eigenvalues $\{E_i\}$ of $H_0$, and
$\tilde{H}_{\rm mod}$ is the matrix representation of the operator in the eigenbasis of $H_0$.
In numerical experiments with different periodically-driven nonlinear potentials, we found that this choice of the basis
guarantees stable performance for $N > 10^3$. Because of the diagonal form of the matrix $H_0$,
it  also simplifies calculation of the nested commutators.

The algorithm can be described as the propagation of the $N \times N$  identity (in the eigenbasis of $H_0$)
matrix over the time interval $T$.
The propagation is realized with a chain of $M$ single-step Magnus propagators over the time interval $h = T/M$,
by performing $L$ Chebyshev iterations (cf. Eqs.~(\ref{chebyshev1}),~(\ref{chebyshev2}))
with the rescaled operator $\tilde{\Omega}(t_{k-1},t_k)$, $k=1,...,M$,  Eq.~(\ref{magnusApp}.
Note, that in order to apply the rescaling procedure  (\ref{newOmega}),
$\Omega \mapsto \tilde{\Omega}$, one has to estimate the
extreme eigenvalues $E_{\rm min}$ and $E_{\rm max}$ beforehand.
We diagonalize the matrix  $\tilde{\Omega}$ at five equidistant time instants $t_j \in [0,T]$,
and use the maximal and minimal values from the collected eigenvalue set
as $E_{\rm min}$ and $E_{\rm max}$. Once the propagation stage is completed,
the result,i.e.  the $N \times N$ unitary matrix $U_T$ is diagonalized
and its eigenvalues $\{\epsilon_{\mu}\}$  and eigenvectors $\{|\phi_{\mu} (0)\rangle\}$,
are written into the output file. An additional propagation round can be performed, now with
eigenvectors $\{|\phi_{\mu} (0)\}$ as initial vectors, in order to calculate
relevant characteristics of the Floquet states.
For example, it can be  the expectation value of a relevant operator $A(t)$, averaged over the one
period
\begin{eqnarray}
\langle A_{\mu}\rangle_T = \frac{1}{T} \int_0^T \langle \phi_{\mu} (t)|A(t)|\phi_{\mu} (t)\rangle dt.
\label{average}
\end{eqnarray}


\section{Models} \label{models}
To test the algorithm, we employed two specific physical cases for the Hamiltonians entering the setup
given by equation ~(\ref{Ham1}).

The first system  is a synthesized model, with the Hamiltonians $H_0$ and $H_{\mathrm{mod}}$ being members
of a Gaussian orthogonal ensemble $\mathrm{GOE}(N)$ \cite{RM} of a variance  $\sigma$, that is a parameter of the system.
Random matrix theory and the corresponding models remain at the center of research in
many areas of quantum  physics \cite{RMphysics}, but it is only  very recently that these two
hitherto disentangled research fields
started to interact  ~\cite{lazarides2, rigol}.

Our second test model consists of a driven $N$-particle Bose-Hubbard dimer \cite{dimer}, with the Hamiltonians
\begin{align}
\begin{aligned}
& H_0 = -\upsilon (\hat{a}_1^\dagger\hat{a}_2 + \hat{a}_1\hat{a}_2^\dagger) + \frac{U}{2} (\hat{n}_1
- \hat{n}_2)^2, ~~~~~~~~~~~~~~~~~~~ \\
& H_{\mathrm{mod}} = (\hat{n}_2 - \hat{n}_1),
\label{Ham_dimer}
\end{aligned}
\end{align}
where $\hat{a}_j^\dagger$ ($\hat{a}_j$) and $\hat{n}_1 = \hat{a}_j^\dagger\hat{a}_j$
are the bosonic creation (annihilation) and particle number operators for the $j$-th site, respectively.
Parameters $\upsilon$ and $U$ are the hopping rate and one-site interaction strength. In the Fock basis
the Hamiltonian $H_0$ acquires a tridiagonal structures,
while $H_{\mathrm{mod}}$ becomes a  diagonal matrix. This model is extensively used in many-body quantum physics, both
in theoretical and experimental domains; e.g., see Ref. \cite{dimer1}.

\section{Implementation of the algorithm on a cluster}
\label{realization}
We now describe a program realization of the algorithm and
its consecutive realization on a high-performance cluster.
Our   $\mathrm{C}$ code employs Intel$^\circledR$ Parallel Studio XE package \cite{IPSXE}.
The main data structures are complex double-precision matrices.
Computational load is distributed among cluster nodes by the standard
Message Passing Interface (MPI).
On each node computationally intensive operations are implemented by
calling BLAS functions from Intel$^\circledR$ Math Kernel Library (Intel MKL),
in shared-memory parallel mode \cite{IMKL}.

\begin{figure}[t]
\begin{tabular}{cc}
\includegraphics[width=0.5\textwidth]{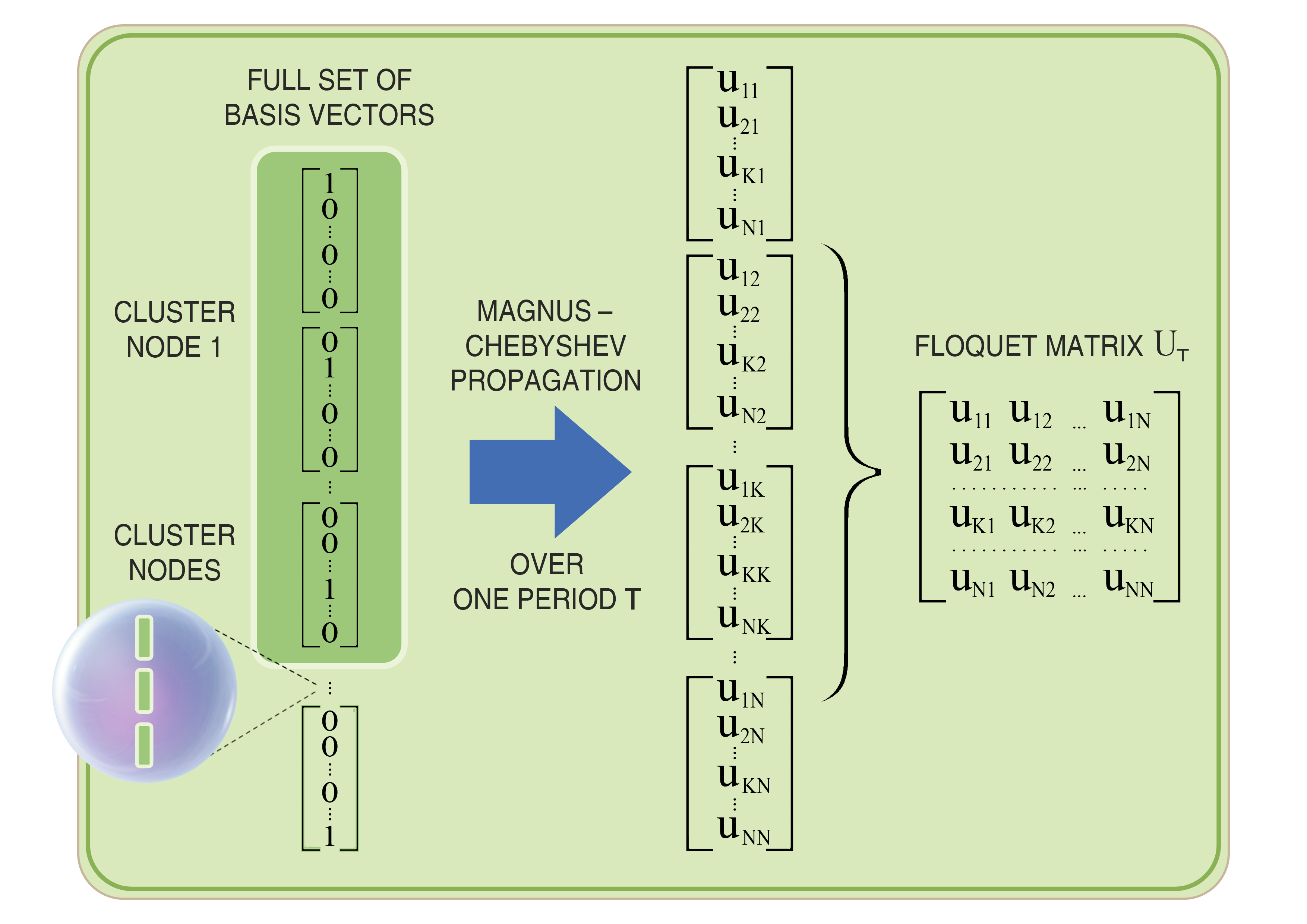}
\end{tabular}
\caption{(color online) Parallel computation of a Floquet operator $U_T$.
The initial  $N\times N$ identity matrix $\dblone$ is sliced into $P$ rectangular
sub-matrices $X_i$, $i = 1,\ldots,P$,
each consisting of $N/P$ basis vectors.
The sub-matrices are then independently propagated on $P$ cluster nodes,
by using the Magnus-Chebyshev propagation algorithm.
The output vectors form the
corresponding columns of the Floquet matrix.}
\label{Fig:2}
\end{figure}

\begin{algorithm*}[htb]
\caption{}
\label{alg:AllBodies}
\begin{algorithmic}[1]
\State initialization, memory allocation
\State \textbf{upload} {system $\&$ method parameters: the number of steps $M$ per period $T$, the number of Chebyshev iterations $L$}
\State calculate basis of the stationary Hamiltonian $H_0$, auxiliary matrices ${\rm diag} E_i$ and $\tilde{H}_{\rm mod}$
\State calculate Bessel functions $J_l(R)$, $R=R(h)$, $l=0, ... ,L$
\State distribute the initial sub-matrices $X_i$, $i=1,...,P$ and precalculated data between $P$ nodes
\For{$k = 1$ to $M$} \Comment{\textit{MPI distributed computational loop}}
\State calculate $\Omega[t_{k-1},t_k]$ for the current $t_k = k h$, $h=T/M$
\State rescale $\Omega \mapsto \tilde{\Omega}$
\State perform $L$ Chebyshev iterations with $\tilde{\Omega}$ and update $X_i$
\EndFor
\State combine $X_i$ into  $U_T$
\State diagonalize $U_T$
\State \textbf{save} eigenvectors and eigenvalues of $U_T$
\State \textbf{release} memory
\end{algorithmic}
\end{algorithm*}

The code consists of three main steps (they are summarized in the
pseudocode shown in Algorithm~\ref{alg:AllBodies}).
In the first step, the program initializes MPI, allocates memory, reads the seed data
and parameters  from configuration files, and makes necessary pre-calculations before
launching the main cycle: calculates the eigenbasis of the Hamiltonian $H_0$ and
auxiliary matrices ${\rm diag} E_i$, $\tilde{H}_{\rm mod}$ (see Eq.~\ref{Ham2}),
the Bessel functions $J_l(R)$, $R=R(h)$, $l = 0, ... ,L$\footnote{The Bessel functions were 
numerically computed using Fortran intrinsic function \textit{bessel\_ jn},}, needed for
the Chebyshev series (see Eq.~(\ref{chebyshev2})),
and nine commutators needed for the Magnus expansion (see Eq.~\ref{magnusApp}).
These computations are performed on each cluster node.
It is important to choose appropriate operational presentation of the
$N \times N$ matrices, starting from the initial identity matrix $\dblone$.
The most straightforward solution is to split $\dblone$ into $N$ vectors,
store the vectors as independent  arrays, and then propagate them
independently and in parallel.
A more efficient solution is to form sub-matrices of initial vectors that allows for parallel
propagation and then make use of the third-level BLAS operations, in particular, matrix-matrix product,
instead of a series of matrix-vector products. As a result, the memory hierarchy would be used in a
more efficient way and a substantial decrease of the computation time would be achieved.

The second step involves the propagation of the initial matrix $\dblone$ over one period $T$.
Because the process is iterative, a parallelization in time is not possible.
However, a data parallelization  is feasible. The initial identity matrix can be
split into $P$ sub-matrices $X_i$, $i = 1, ... ,P$, each consisting of $N/P$ basis vectors,
which are then distributed among $P$ cluster nodes.
Therefore, the first $N/P$ rows are propagated on the first node,
the next $N/P$ rows on the second node, etc. (according to the C row-major order, initial vectors are written as rows).  This idea is sketched in Fig.~\ref{Fig:2}. The scheme possesses a potential minor drawback that could be encountered in the case of a large number of processing units, when splitting could cause a strong imbalance in
This might affect the performance of the mathematical kernels,
which were not developed to handle ``thin'' matrices consisting of a few rows and thus limits
number of processing units that could be used to accelerate the propagation.
The major advantage of the scheme, however, is a next to uniform distribution of the workload among the
nodes. Together with a constant number of operations on each step, this  allows
to estimate the scaling of the overall computing time with $P$.

\begin{table*}[htb]
\caption{Single-node performance: Execution times (in \textit{sec}) as a function of
system size $N$. Multi-threaded version of the code employs all $16$ node's cores on shared memory.
Columns ii-iv and vi present  data obtained for $M = 10^2$ time steps per period and
$L = 50$ Chebyshev iterations on every step. To get an estimate for $M = 10^4$, the time needed
to calculate  $\Omega$ and perform Chebyshev iterations were extrapolated (last column),
see text for more details.}
\label{table:1}
\vspace{6pt}
\small
\centering
\begin{tabular}{c c c c c c c}
\hline
System size, & Auxiliary 				& Time of &	Chebyshev 		& Diagonalization &	Total time		& Total	time															 \\
N	& computations time &	$\Omega$ calculation  &	iterations time & time 			&			 $M = 10^2$	 \					& $M = 10^4$, extrapolation \\
\hline		
\	256 	\		&	0.2 	& 0.4			& 4.3				& 0.2			& 5.1      	& 470.4 		   \\
\ 512 	\ 	&	0.4		& 1.8			& 25.3			& 0.6			& 28.1     	& 2 711.0 	   \\
\ 768 	\		& 1.3		& 3.2			& 79.9			& 1.4			& 85.8     	& 8 312.7 	   \\
\ 1 024 \ 	&	1.8 	& 8.3			& 177.3			& 2.6			& 190.0   	& 18 564.4 	   \\
\ 1 536	\ 	& 4.3		& 18.9		& 559.1			& 7.3			& 589.6    	& 57 811.6     \\
\ 2 048	\ 	& 8.6		& 33.2		& 1 296.6		& 16.0		& 1 354.4  	& 133 004.6    \\
\ 3 072	\ 	& 23.5	& 72.1		& 4 179.2		& 51.0		& 4 325.8	 	& 425 204.5    \\
\ 4 096	\ 	& 49.2	& 126.0		& 9 730.5		& 117.5		& 10 023.2 	& 985 816.7    \\
\ 5 120	\ 	& 100.5	& 184.3		& 18 667.2	& 242.1		& 19 194.1 	& 1 885 492.6  \\
\ 10 240\ 	& 755.3	& 919.7		& 181 722.3	& 1 857.7	& 185 254.9 & 18 266 809.4 \\
\hline
\end{tabular}
\end{table*}

A single propagation step realizes the recipe given at the end of Section~\ref{algorithm}.
By employing MKL functions, we calculate the matrix $\Omega(t_{k-1},t_k)$ following the Magnus expansion (\ref{magnusApp}). It is computed independently on each cluster node, as the small computing time does not justify parallelization on a distributed memory.

The computationally intensive part of the
algorithm is the approximation of the action of the matrix exponent
by Chebyshev's iterations, Eqs.~(\ref{chebyshev1},\ref{chebyshev2}),
and the further updating of propagated sub-matrices on each cluster node.
The mathematical core of this step is the multiplication of complex double-precision dense matrices
(it is realized with  \textit{zgemm} routine  \cite{MKL_docs}). This part of the algorithm is fully parallel.

During the final, third step the program assembles sub-matrices into the Floquet matrix
and diagonalizes the latter by using a multi-threaded Intel MKL implementation (we use \textit{zgeev}
routine \cite{MKL_docs}). For the
matrix size $N \sim 10^4$, a  multi-core implementation is sufficient. Finally, the
results of the diagonalization are written to the output files, the memory is deallocated,
and MPI is finalized.

\section{Program performance and scalability analysis} \label{single_node}

In this section we present the performance analysis of the code.
Test runs were performed on the ``Lobachevsky''
supercomputer at the Lobachevsky State University of Nizhni Novgorod \cite{lobachevsky}.
We employed up to $64$ computational nodes, with the following configuration per node:
$2 \times$ Intel Xeon E$5-2660$ CPU ($8$ cores, $2.2$ GHz), $64$ GB RAM, OS Windows
HPC Server $2008$. We  use  Intel MKL, Intel $\mathrm{C/C++}$ Compiler, and
Intel MPI from Intel Parallel Studio XE \cite{IPSXE}.
All   parallel versions of computationally
intensive routines from MKL utilized 16 cores on each node.

To test the performance of the program  we use the synthesized random-matrix model as a benchmark (see
Section~\ref{models}). In this case, Hamiltonians $H_0$ and $\tilde{H}_{\mathrm{mod}}$ in Eq.~(\ref{Ham2})
were generated randomly from
the $\mathrm{GOE}(N)$ ensemble of the unit variance $\sigma = 1$ \cite{RMgen}.
The driving function is $f(t) = \cos(\omega t)$ with $\omega = \pi$ \cite{effective}.


\textit{Single-node performance.} To test a single-mode performance, we use $L = 50$
Chebyshev iterations on every step.
The number of steps per period, $M = 10^2$, was
used for testing the program.
The execution time for larger values of $M$
can be easily extrapolated: due to the linear increase of operations number with iterations in time,
it is sufficient to find and appropriately scale execution time of the core part
of the code, and add execution time of the other parts, which are independent of $M$.
Table~\ref{table:1} presents the dependence of the execution time on the size of the model system.
The last column of Table~\ref{table:1} presents estimates
for the case when the number of steps is increased $100$-fold, i.e. for $M = 10^4$.

To gain a further insight, we analyze a single-node performance in some more detail.
We consider two metrics, both normalized by the number of operations, $\mbox{OC}_N$, required to make calculations for a system
of a size $N$.
Namely, we calculate the operation rate $R$б that is the number of operations per unit
time, and take the value obtained for $N = 256$ as a unit measure. The first metrics reads
$R_N = (\mbox{OC}_N / \mbox{OC}_{256}) / (\mbox{TIME}_N / \mbox{TIME}_{256})$,
where $\mbox{TIME}_N$ is the execution time of the code for the system of size $N$.
Further, we consider a similar quantity, $P_N$, where the number of
operations is estimated by $N^3$,
according to the scaling of the most computationally intensive and most frequently
called MKL subroutine \textit{zgemm}. Table~\ref{table:2}
presents $R_N$ and $P_N$ as functions of $N$.

\begin{table}[htb]
\caption{Single-node performance. Computational
intensity and efficiency measures, $R_N$ and $P_N$, as functions of the model system size $N$.
Multi-threaded implementation of the algorithm  on a single cluster node ($16$ cores on shared memory)
was  used. The number of steps per period is $M = 10^2$ with $L = 50$ Chebyshev iterations on every step.}
\label{table:2}
\vspace{6pt}
\small
\centering
\begin{tabular}{c c c c c}
\hline
System size,	& Total time, & Operations count,  & $R_N$	& $P_N$ \\
N	& TIME$_N$, in \textit{sec} & OC$_N$, in \textit{mln} &	\			& \ 		\\
\hline			
\ 256	\ 	& 5.1				& 173 612				& 1.00	& 1.00 \\
\ 512	\ 	& 28.1			& 1 365 642			& 1.43	& 1.45 \\
\ 768	\ 	& 85.8			& 4 594 916			& 1.57	& 1.60 \\
\ 1 024	\ & 190.0			& 10 875 688		& 1.68	& 1.72 \\
\ 1 536	\ & 589.6			& 36 655 208		& 1.83	& 1.87 \\
\ 2 048	\ & 1 354.4		& 86 961 706		& 1.89	& 1.93 \\
\ 3 072	\ & 4 325.8		& 292 683 840		& 1.99	& 2.04 \\
\ 4 096	\ & 10 023.2	& 693 473 512		& 2.03	& 2.08 \\
\ 5 120	\ & 19 194.1	& 1 354 053 950	& 2.07	& 2.13 \\
\hline
\end{tabular}	
\end{table}

The number of operations is calculated by Intel$^\circledR$ VTune$^{\rm TM}$
Amplifier profiler \cite{IVTA} and returned to
CPU performance counter $\mbox{SIMD}\_\mbox{FP}\_\mbox{256.PACKED}\_\mbox{DOUBLE}$ \cite{IVTAUG}.
This variable contains the number of issued Advanced Vector Extensions (AVX) instructions
for processing double precision values \cite{AVX}.
The choice of this counter is based on the fact that almost all computations in
our code occur in Intel MKL BLAS routines.
Note, however, that this estimate of the number of operations is not exact.
It is well known that for the current architectures the profiler tends to overestimate
this number, since it counts the number of instructions issued but retired.
Nevertheless, this estimate is reliable for CPU-bound processes.

\begin{figure}[htb]
\begin{center}
\includegraphics[width=\columnwidth,keepaspectratio,clip]{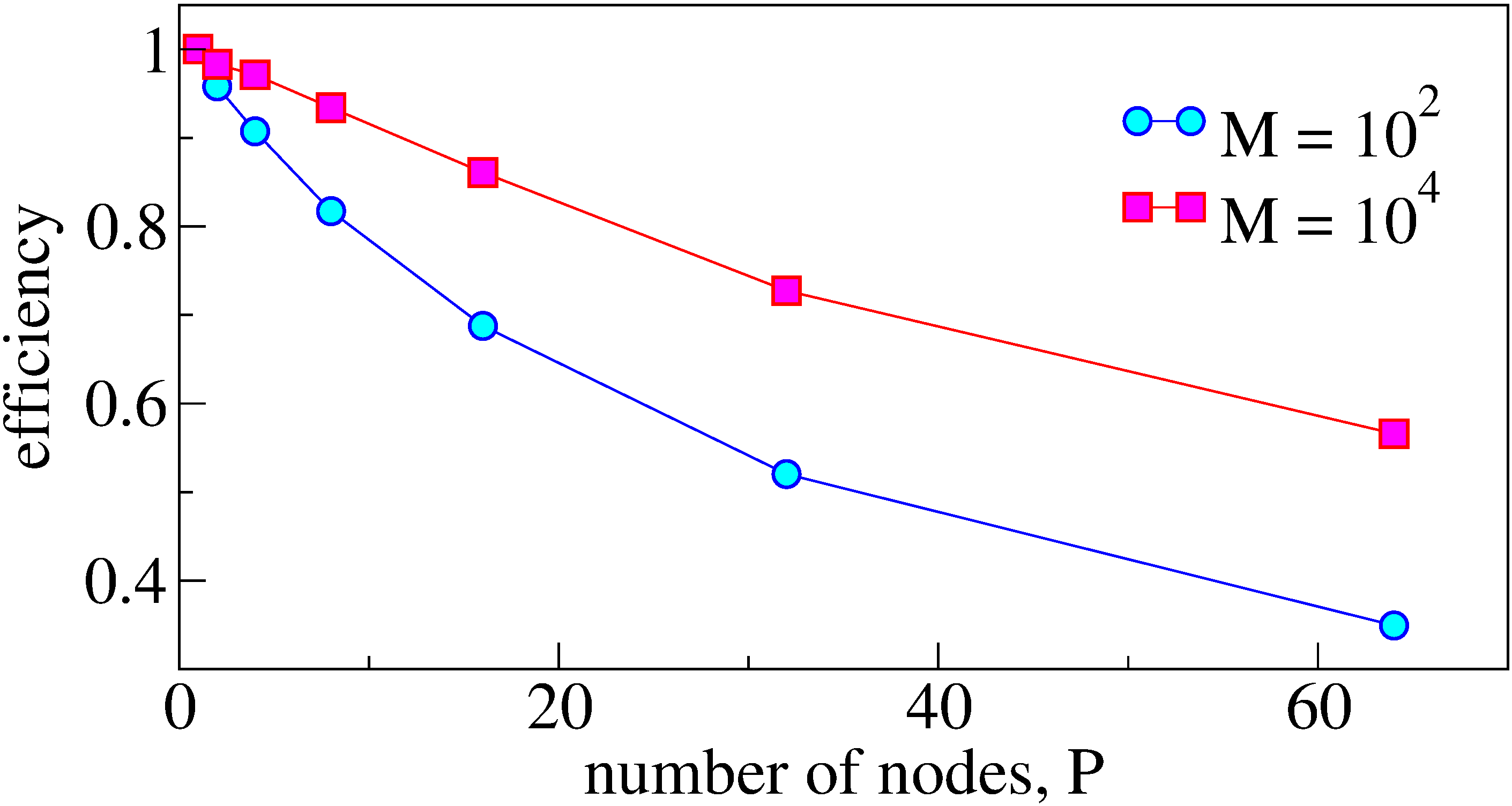}
\caption{(color online) Computational efficiency as a function of number of cluster nodes $P$.
Results are shown for
two values of number of steps per period, $M = 10^2$ and $M = 10^4$. The parameters are $N = 5~120$ and $L = 50$.}
\label{figure:2}
\end{center}
\end{figure}

The behavior of $R$ and $P$ as  functions of $N$
are presented in two last columns of Table~\ref{table:2}.
The efficiency increases with
the size of the model system,
doubling  for $N = 5~120$ as compared to $N = 256$.
That is because of the increasing efficiency in evaluation of larger matrices
of the BLAS computational kernels in the parallel regime.
While the efficiency grows with the system size,
the execution time also increases, mainly because the increase of the number of steps per period needed,
and for  $N = 5~120$  the estimated calculation time is about $22$ days.
Therefore, a  multi-node parallelization is required to decrease
the calculation time to more realistic time scales.

\begin{table*}[htb]
\caption{Scalability analysis: execution times (in \textit{sec}) of
different steps of the algorithm and speed-up factors (in \%) are shown as
functions of the number of employed cluster nodes. Each node runs
multi-threaded version on $16$ cores on shared memory.
The parameters are $N=5~120$, $L = 50$.
}
\label{table3}
\vspace{6pt}
\small
\centering
\begin{tabular}{c c c c c c c c c}
\hline
Number	 & Auxiliary 	  & Time of $\Omega$ &	Chebyshev & Chebyshev 	& Total 		 & Speed up 		&	Total				& Speed up 		\\
of nodes,& computations &	calculation      & iterations & iterations	& time		 	 & $M = 10^2$   &	time        & $M = 10^4$  \\
P				 & 		time		  &	\				         & time 			& speed up 		& $M = 10^2$ &	\ 					&	$M = 10^4$	& \						 \\
\hline			
\ 1	\   & 100.5	& 184.3	& 18 667.2	& 1.0  & 19 194.1		& 1.0 	& 1 885 457.9 & 1.0  \\
\ 2	\   & 186.4	& 181.5	& 9 406.9	  & 2.0  & 10 017.8		& 1.9 	& 959 150.8		& 2.0  \\
\ 4	\   & 195.3	& 180.2	& 4 670.7	  & 4.0  & 5 288.8		& 3.6 	& 485 400.8		& 3.9  \\
\ 8	\   & 173.6	& 178.7	& 2 341.2	  & 8.0  & 2 935.8		& 6.5 	& 252 305.2		& 7.5  \\
\ 16 \	& 137.1	& 178.4	& 1 187.2   & 15.7 & 1 744.7 		& 11.0 	& 136 882.6		& 13.8 \\
\ 32 \	& 103.9	& 178.7	& 628.0			& 29.7 & 1 152.6		& 16.7 	&	81 006.1		& 23.3 \\
\ 64 \	& 98.9	& 178.6	& 338.3			& 55.2 & 858.6			& 22.4 	& 52 053.6		& 36.2 \\
\hline
\end{tabular}
\end{table*}


\textit{Strong scalability of the algorithm}. We next analyze the performance of the algorithm on a cluster.
To benchmark the code, we  use the random-matrix model of the size  $N = 5~120$
and launch the code on $P = \{1, 2,..., 2^j,...,2^6\}$ cluster nodes, using the
multi-threaded implementation on each node as before. The results are
summarized in Table~\ref{table3}. Let us note that the time needed for the diagonalization
of  $N = 5~120$ matrix is $242.1$ \textit{sec} and does not
depend on $P$ (see column vi in Table~\ref{table:1}).
Therefore, it is omitted from the further analysis.

%

For $M = 10^2$ the code  accelerates as the number of nodes
increases to $64$ ($1~024$ computational cores in total),
though  the efficiency of parallelization,
defined as the ratio between the speed up and the number of nodes,
drops to $35\%$.
That is because for this, relatively small, number of integration steps
the execution time of serial parts of the code becomes
comparable to that of the parallel code (Chebyshev iterations), which scaling efficiency,
in turn, is about $86\%$.
Taking into account that practically reasonable
computations  require much larger
numbers of integration steps, it is expected that the
efficiency of the code will increase with $N$. It is confirmed by the results of test runs, see last two columns
of  Table~\ref{table3}. In particular,
for $N = 5~120$ and $M = 10^4$ the code is executed on $64$ cluster nodes
in  $14.5$ hours,  demonstrating the efficiency of parallelizing about $57\%$.
Figure~\ref{figure:2} shows the dependence of the efficiency on the number of employed cluster nodes.

\section{Applications}
\label{results}

In this section we test the accuracy of the algorithm by using
two physical model systems which we described in Section \ref{models} above.

\begin{figure*}[htb]
\includegraphics[width=0.99\textwidth]{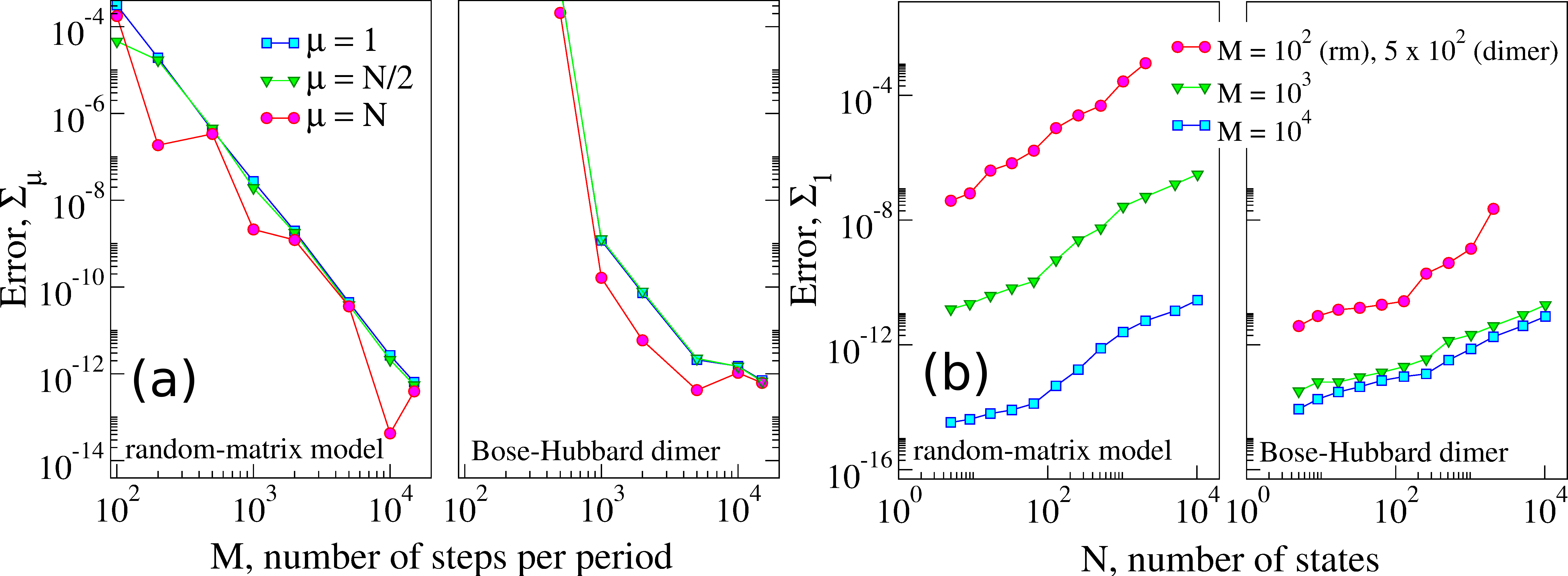}
\caption{(color online) Phase error $\Sigma_\mu$, Eq.~(\ref{error}), as a function of (a)  number of steps
per period $M$ ($N = 1025$)
and (b) size $N$ of the model system. (a) The dependence $\Sigma_\mu$  for the three Floquet
states $|\phi_{\mu} \rangle $: ($\square$) groundstate, $\mu = 1$; ($\triangledown$) the  state from the center of the energy spectrum, $\mu = N/2$;
($\circ$) the highest-energy state, $\mu = N$. (b) The dependence $\Sigma_1$  for the  Floquet groundstate.}
\label{Fig:3}
\end{figure*}

The random-matrix model was already specified in Section~\ref{single_node}.
For the dimer model, Eq.~(\ref{Ham_dimer}), we use  parameters $\upsilon = 1$ and $U = U'\cdot N = 2$.
The one-site interaction is scaled with the number of bosons, $N-1$, to match in the limit $N  \rightarrow \infty$
the classical mean-field Hamiltonian \cite{dimer1,dimer2},
\begin{eqnarray}
H_{\mathrm{cl}}(z,\nu) = \frac{U'}{2}z^2 - 2\upsilon \sqrt{1 - z^2}\cos(\nu) + 2z\cdot f(t).
\label{mean-field}
\end{eqnarray}
The mean-field variables $z$ and $\nu$
measure the population imbalance and relative phase between the dimer sites, respectively.
The driving function $f(t) = f_{\mathrm{dc}} + f_{\mathrm{ac}}
\cos(\omega t)$ consists of two components, a constant dc-bias $f_{\mathrm{dc}} = 2.7$ and
single-harmonic term $f_{\mathrm{ac}} \cos(\omega t)$, with the amplitude $f_{\mathrm{ac}} = 2.5$ and frequency
$\omega =3$. We  use the phase space of the mean-field system, Eq.~(\ref{mean-field})
together with the semi-classical eigenfunction hypothesis \cite{hypoth} and the
concept of ``hierarchical eigenstates'' \cite{hier},  for a ``quantum'' benchmarking of the program.

\begin{figure*}[t]
\includegraphics[width=0.99\textwidth]{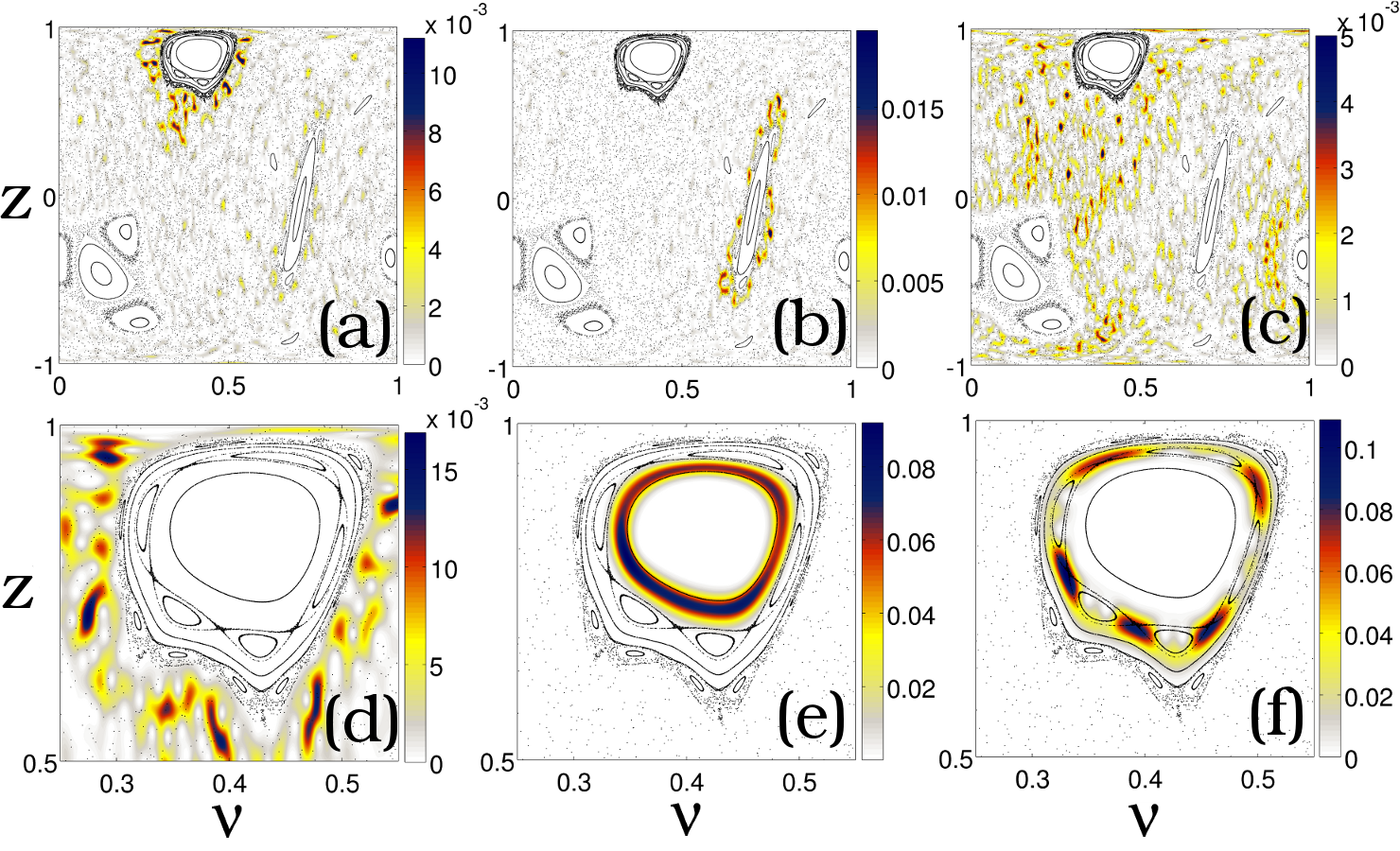}
\caption{(color online) Husimi distributions for hierarchical, chaotic and regular Floquet states
of the dimer model with $N = 2~499$
bosons. Dots show the Poincar\'{e} section for the  mean-filed Hamiltonian, Eq.~(\ref{mean-field}).
Circle-like formations correspond to the KAM tori \cite{licht}, the solutions of the mean-field
system inside regular islands. (a,b,d,)
Hierarchical Floquet states of the quantum system:  $\mu = 118$(a),  $\mu = 1~024$ (b), $\mu = 101$ (d, zoom).
(c) Chaotic Floquet state, $\mu = 1002$. (f,g) Two regular Floquet states: $\mu = 30$ (f) and $\mu = 55$ (g).}
\label{Fig:4}
\end{figure*}

Once the diagonalization  of $U_T$ is completed, the program
initiates an additional round of $T$-propagation to calculate the average energies of the Floquet states
with $A(t) = H(t)$ in Eq.~(\ref{average}). Finally, the Floquet states are
sorted in ascending order according to their average energies.
To quantify the accuracy, we adapt the idea of overlap phase error ~\cite{talezer} and modified
it to account for the periodicity of the Floquet states,
\begin{eqnarray}
\Sigma_{\mu} = \lvert \dblone- \langle \phi_{\mu}(0) |U_T| \tilde{\phi}_{\mu}(0) \rangle \rvert.
\label{error}
\end{eqnarray}
where $\tilde{\phi}_{\mu}(0)$ is the Floquet state calculated with
the time step $\tilde{h} = h/2$. Note that the error is state-specific.

Figure~\ref{Fig:3} presents the error $\Sigma_{\mu}$ as a function of number of steps per period $M$
and number of system states $N$.
A linear dependence of $\log_{10}\Sigma_{\mu}$ on $\log_{10}M$ observed for the random-matrix
model is typical for stepwise integrators.
The error convergence with the number of steps does not saturate up to largest value
$M = 10~240$. In the case of the dimer, however, the error does not reveal
the power-law  scaling and demonstrate  a noticeable  saturation.
The difference in the scalings can be attributed to the differences in the spectral properties of the
matrices $H_0$ and $H_{\mathrm{mod}}$: while in case of the random-matrix model the level spacings of the Hamiltonians
are characterized by a probability density function (pdf) with a gap near zero,
the level spacings in the energy spectrum of the integrable  dimer Hamiltonian are characterized by a gapless
Poisson pdf \cite{qchaos}. The Floquet groundstate $|\phi_{1} \rangle$ turns to
be the most sensitive to the discretization of the unitary evolution (see Fig.~\ref{Fig:3}(a)).
We use this state to test dependence of the error on the size of the model system.
Figure~\ref{Fig:3}(b) shows that the scaling of
$\Sigma_{\mu}$ with $N$ is qualitatively similar for both models.

Now we turn to the quantum benchmarking of the algorithm with the dimer model. Following the
semi-classical eigenfunction hypothesis \cite{hypoth,qchaos}, the Floquet states of the
model in the limit $N \gg 1$ can be sorted according  to the location of the state's Husimi (or Wigner)
distributions \cite{qchaos} in the mean-field phase space.
The latter is \textit{mixed} in the case of the driven Hamiltonian (\ref{mean-field}), so that
regular and chaotic regions coexists  \cite{licht} (see dots on the Poincar\'{e} sections on Fig.~\ref{Fig:4}).
If the distribution of a Floquet state
is localized inside a regular region the state is labeled  ``regular''. When the distribution
is located in the bulk of
the chaotic sea, the corresponding Floquet states is ``chaotic''.  The regular regions, called ``islands'',
are often organized in a
hierarchical way, forming fine island-near-island structures. By increasing the number of bosons in the dimer,
it is possible to resolve these structures with  higher level of detail. It is important to understand, however,
that the gradual motion towards the semi-classical limit does not simplify the quantum problem. On the contrary,
the increase of $N$ increases the size of the Hamiltonian matrices, the number of the basis vectors,
and, what is most dramatic, decreases the mean spacing between
quasienergies $\epsilon_{\mu}$ (which for any $N$
remain restricted to the fundamental stripe $[-\hbar \omega/2, \hbar \omega/2]$). Therefore, the accuracy
of calculations has to be increased in parallel, otherwise the numerically obtained Floquet states would represent
incoherent mixtures of actual Floquet states (while an error in the unitarity of the propagation
may remain reasonable small). A theoretical interpretation of so obtained numerical results
might lead to wrong conclusions.

The accuracy of the scheme can be checked by using the chaotic - regular dichotomy. Figures 4 (e,f) show
Husimi distributions  for two regular Floquet states, while Fig.~4 (c) presents the distribution for
a chaotic state. The accuracy can be tested even more carefully with
a third class of quantum  states, located on the interface
between chaotic and regular ones. These \textit{hierarchical} \cite{hier} states  are supported
by the classical phase-space  structures located on the chaotic sea's offshore around regular  islands.
The Husimi distribution of a hierarchical state is
concentrated in the immediate  exterior of the corresponding island.
Hierarchical states are exceptional in the sense that
their absolute number increases sub-linearly with the number of states, $N_{hier} \sim N^{\chi}$, $\chi < 1$,
so that their relative fraction $N_{hier}/N$ goes to zero in the limit $N \rightarrow \infty$ \cite{hier}.
These states must be carefully  selected from the complete set of $N$ Floquet states.

Hierarchical states are coherent superpositions of  many basis vectors and  therefore sensitive
to the phase errors. Even a small mismatch  in vector phases blurs the interference pattern and causes the
flooding of the state's Husimi distribution into the island. The high coherence of the superpositions
is also a trait of the regular Floquet states but there is an important difference:Quasienergies $\epsilon_{\mu}$
of the hierarchical states  are
randomly distributed  over the interval $[-\hbar \omega/2, \hbar \omega/2]$ while the quasienergies of regular
and chaotic states tend to cluster in different regions \cite{hier}. Because of that, phases of
hierarchical states are most vulnerable to the error produced by the numerical propagation.
We selected several hierarchical states for the dimer model with $N = 2~499$\footnote{We use the definition of the Husimi distribution for the dimer given in Refs.~\cite{husimi1}.
The expression involves summation over the series of square roots of binomail coeffcients of the order $N$.
We did not find an alternative expression which allows to avoid term-by-term summation.
Although we calculated the Floquet states for the dimer
with $10~239$ bosons, we were not able to go beyond the limit $N = 2~500$ when calculating
Husimi distributions.} bosons and inspect
their Husimi distributions \cite{husimi2}, see Figs.~4 (a,b,d).
The offshore localization and absence of the flooding [see  zoomed distribution on Fig.~4 (d)] are clearly visible.

\section{Summary and Outlook}
\label{summary}

We have put forward a method to calculate  Floquet states of
periodically-modulated quantum system with  $N \geq 10^4$ states. Our method is advantageous in that it is
scalable and therefore  well suited for its implementation on parallel
computers. Our study uses massively parallel clusters as efficient devices to explore
complex quantum systems far from equilibrium, thus
answering the need of several, actively developing, research
fields involving quantum physics \cite{bukov,eisert,top,majorana,gauge,lazarides,lazarides2,rigol,abanin}.

The method particularly allows for  improvements, such as the increase of the order of the Magnus expansion \cite{blanes} or
the use of commutator-free Magnus approximations \cite{alverman}. With respect to further acceleration of the code
for systems with $N \leq 10^4$ states, there is a promising perspective  related to
the fact that the main contribution
to the computation time stems from the BLAS operations. These operations fit GPU and Intel Xeon Phi
architectures very well. By our estimates, even a straightforward  implementation of the most
computationally intensive  Chebyshev iteration stage on a heterogeneous CPU+GPU configuration will result
in a three-fold speedup. A yet further speed-up can be obtained by using multiple accelerators.

There are several interesting research directions
for which the proposed algorithm may serve as a useful starting point.
For example, there is the perspective
to resolve Floquet states of even larger systems by
applying the spectral transformation  Lanczos algorithm \cite{spectral} to the corresponding
time-\textit{independent} super-Hamiltonians, to name but a few.
Because the  super-Hamiltonian elements can be generated on
the fly, this idea potentially would allow to calculate Floquet states of a system
with $N \sim 10^5$ states for $F \sim 10^4$ Fourier harmonics, by employing massively parallel exact diagonalization
schemes \cite{exact}. Note, however, that the eigenvalues of the Hamiltonian super-matrix
(as well as the respective quasienergies) are merely phase factors and are not directly related with the properties
of the corresponding Floquet states (average energy, etc.). Therefore, some targeting of the algorithm
to the low-energy states is required. Our method can be used to locate
the relevant Floquet eigenvectors in the quasi-energy spectrum of a system with a smaller number of states;
combining this with a knowledge of the spectrum scaling  with $N$, one can target
the Lanczos algorithm.

Another direction relates to the computational physics of \textit{open} periodically-modulated 
quantum systems that interact with a large environment (heat bath).
Asymptotic states of such systems are 
affected by the combined effects  of modulation and the decoherence induced by  the
environment \cite{brower}. Due to linearity of the model equations describing the evolution of 
the density matrices of the systems, the corresponding asymptotic states are 
specified by time-periodic density matrices, which can be called ``auntum attractors''.
There is presently   limited knowledge
about the theme of quantum attractors beyond the limit of the rotating-wave approximation  \cite{petr}.
In the Floquet framework, the attractor's density matrix is a zero-eigenvector of
the corresponding non-unitary Floquet super-operator, which acts in the space of $N \times N$ Hermitian matrices.
This super-operator can be constructed  by propagating
the identity operator -- but now in the space  of $N \times N$ matrices.
The propagation stage can be realized
by using Pad\'{e} approximation \cite{exp} or with Newton or Faber polynomial schemes \cite{faber},
while the question  whether there exists
a possibility to generalize the Magnus expansion to dissipative quantum evolution equations remains open.
As the number of the basis matrices scales as  $N^2$, the
scalability of  non-unitary propagation algorithms then presents an even more demanding task.

\section{Acknowledgments}\label{acknowledgment}

The authors acknowledge support of Ministry of Education and Science of the Russian Federation, Research Assignment No. 1.115.2014/K (Sections 2--6) and the Russian Science Foundation grant No. 15-12-20029 (Section 7).


\begin{thebibliography}{1}
\bibitem{effective} We set
the effective Planck constant to $\hbar = 1$ for both models.
\end{thebibliography}
\end{document}